\RequirePackage{fix-cm}
\documentclass[smallextended]{svjour3}
\journalname{}
\renewcommand{\makeheadbox}{}
\usepackage{mathptmx}
\usepackage{graphicx}

\usepackage[utf8]{inputenc}
\usepackage[T1]{fontenc}
\usepackage{amsmath,amssymb}
\usepackage{booktabs}
\usepackage{float}
\usepackage{array}
\usepackage{microtype}
\usepackage[round,authoryear]{natbib}
\usepackage{hyperref}
\usepackage{xurl}
\usepackage{graphicx}

\newsavebox{\audittablebox}
\let\originaltabular\tabular
\let\endoriginaltabular\endtabular
\renewenvironment{tabular}[1]{\begin{lrbox}{\audittablebox}\originaltabular{#1}}{\endoriginaltabular\end{lrbox}\typeout{AUDIT-TABLE-WIDTH: \the\wd\audittablebox; LINEWIDTH: \the\linewidth}\usebox{\audittablebox}}

\hypersetup{
  pdftitle={From Candidate Admission to Task-Level Return: Auditing LLM-Generated Code under Execution Contracts},
  pdfauthor={Hiroyasu Usami, Keisuke Hara, Ayato Tsuboi, Naohiko Matsuda},
  pdfsubject={Execution contracts, candidate admission, and task-level return},
  pdfkeywords={code generation, generated tests, execution contracts, candidate admission, task-level selection, reproducibility},
  colorlinks=true,
  linkcolor=blue,
  citecolor=blue,
  urlcolor=blue
}

\title{From Candidate Admission to Task-Level Return:\\
Auditing LLM-Generated Code under Execution Contracts}
\author{Hiroyasu~Usami \and Keisuke~Hara \and Ayato~Tsuboi \and Naohiko~Matsuda}
\authorrunning{Hiroyasu Usami et al.}
\titlerunning{Candidate admission and task-level return}
\institute{H. Usami, K. Hara, A. Tsuboi \at
Department of Computer Science, Graduate School of Engineering, Chubu University, Kasugai, Aichi 487--8501, Japan\\
\email{usami@fsc.chubu.ac.jp}
\and N. Matsuda \at Mitsubishi Heavy Industries, Ltd., Research \& Innovation Center, Heat Transfer Research Department, Takasago, Hyogo 676--8686, Japan}
\date{}

\begin{document}
\maketitle

\begin{abstract}
Admission rules retain generated programs, whereas selection policies return one candidate per task. We audit how these decisions interact with execution contracts in two separately frozen, disjoint HumanEvalPlus cohorts, each comprising 32 tasks, four samples per task per profile and two served profiles. Under the inline execution contract, a structural-and-self-test screen admits 4--9 expanded-suite nonpasses per 80--107 admissions and excludes 14--43 suite passes per 128 planned attempts. A restricted module-supply condition adds 20 and 14 passing admissions for one profile in the respective cohorts, plus one follow-up nonpass, and reduces the excluded suite passes to 14--23. All fixed first-admitted outcomes and abstentions remain unchanged. In the original pool, the added passes enter already all-passing admitted pools; in the follow-up, four of 24 common sample orders lose one passed task. Selected-subset diagnostics separately address registration-sensitive completion and reference-relative assertion disagreement, accounting for input domains and comparator tolerances. In these high-pass pools, candidate retention and task-level return diverge and are reported separately. Results are conditional on the saved pools, contracts and pinned suites; downstream training effects are unmeasured.
\end{abstract}

\section{Introduction}

Admission and selection are different decisions in a generated-code pipeline: an admission rule decides which attempts are retained, and a selector decides which retained candidate, if any, is returned for a task. A generated program can complete its own tests and fail an independent suite; it can also fail its own tests and pass that suite. We provide a complete accounting of candidate admission and task-level return under two execution contracts within the same saved candidate pools.

Executable benchmarks make these distinctions measurable. HumanEval introduced function-level tasks and pass@$k$ estimation \citep{chen2021codex}; MBPP and APPS extend executable evaluation to other task distributions \citep{austin2021mbpp,hendrycks2021apps}. EvalPlus exposes failures missed by original benchmark tests \citep{liu2023evalplus}. CodeT selects candidates through execution agreement on generated tests \citep{chen2022codet}, while KodCode collects co-generated solutions and tests through self-verification and coverage checks \citep{xu2025kodcode}. The evidence practice at issue is the reuse of a screening outcome as a label for retaining generated code.

We examine a fixed structural-and-self-test screen in two separately frozen, disjoint HumanEvalPlus cohorts, each containing 32 tasks, four samples per task per profile, and two served profiles. The suites were fixed before generation and use reference-derived outputs and task-specific comparators. The original cohort is selected without candidate outcomes; the follow-up is motivated by earlier results and frozen before its own generation. The screen is one fully specified completion-based admission rule, not a representative or optimized production filter. Fixing its three-assertion requirement lets us state exactly what its admission label observes, rather than optimize that rule after seeing its outcomes. A data curator needs both the admitted suite nonpasses and the excluded suite passes; a task solver needs the returned pass, nonpass or abstention. This study measures both and leaves downstream training value unmeasured.

We ask three questions:
\begin{enumerate}
  \item How does joint admission disagree with the independent suite in both directions?
  \item Which selected rejections depend on the execution contract, and which self-test expectations disagree with the pinned reference?
  \item How do changed admissions affect task-level returns and abstentions under declared selectors?
\end{enumerate}

For Flash and gpt-oss, admitted expanded-suite nonpasses are 4/80 and 7/100 in the original cohort and 4/83 and 9/107 in the follow-up. The inline screen excludes 43 and 20 passing attempts originally and 34 and 14 in the follow-up, per 128 planned attempts per profile. Under restricted module supply the corresponding exclusions are 23 and 20, and 20 and 14. The prompt does not specify a module name or placement, so inline import failure is distinct from a wrong expected output. Reference substitution, a same-batch registration control, and assertion-local probes diagnose selected original-cohort rejections without extending that diagnosis to the follow-up.

The symmetric inline/module-supply comparison then links retention to return. Restricted module supply adds 20 and 14 passing Flash admissions in the respective cohorts, plus one follow-up nonpass. Every fixed first-admitted suite outcome and abstention remains unchanged. In the original pool, added passes enter already all-passing admitted pools; the unchanged uniform expectation follows from that saturation. In the follow-up, four of 24 common sample orders lose one passed task. These saved-pool contrasts expose what an admission count alone leaves unspecified.

The study contributes three linked results:
\begin{enumerate}
  \item Complete bidirectional accounting of admitted independent-suite nonpasses and excluded passes in two separately frozen cohorts, under both inline and restricted module-supply contracts.
  \item Selected-subset diagnostics of contract-sensitive completion and assertion/reference disagreement through implementation substitution, a same-batch registration control, and input-domain/comparator checks.
  \item Task-level replay with explicit abstention and all 24 common sample orders, explaining when additional passing admissions leave returned-suite outcomes unchanged.
\end{enumerate}

A change of execution contract changes which generated candidates are admitted, but it need not improve the task-level returns or abstentions of a declared selector. This audit traces that relationship within the same saved candidate pools. Supporting historical, oracle-diagnostic, and finite-domain controls remain separate populations; Section~\ref{sec:limitations} states the corresponding claim boundaries.

\section{Related Work}
\label{sec:related}
\paragraph{The oracle problem and generated tests.}
The test-oracle problem concerns how to determine a program's expected behavior. Barr et al. survey oracles derived from specifications, models, contracts, metamorphic relations, and human knowledge \citep{barr2015oracle}. We audit the evidence attached to generated-code admission, rather than propose an oracle. TestPilot generates and repairs JavaScript unit tests from signatures, implementations, documentation examples, and failure feedback \citep{schaefer2024testpilot}. It evaluates test generation and coverage; we compare code-and-self-test admission with a separately pinned reference-relative suite. Assertion count, coverage, and self-consistency are not treated as correctness labels.

TestGen-LLM extends existing human-written test classes and filters additions for buildability, repeated passage, and incremental coverage \citep{alshahwan2024testgen}. MuTAP repairs mismatched assertion expectations using the program under test's actual outputs and uses surviving mutants to guide further test generation \citep{dakhel2023mutap}. These criteria concern improving tests for an existing program. We compare normal completion of jointly generated code and self-tests under their declared execution contract with passage of a separately pinned reference-relative suite. We do not evaluate those tools or claim that co-generated tests are generally ineffective. The observed disagreements instead delimit what a completion-based admission label can support.

KodCode reports 78 of 80 retained MBPP solutions passing independent tests after manual review of format and numeric edge cases \citep[Section~3.1]{xu2025kodcode}. Its generation, coverage, retry and adjudication policy differs from our fixed-contract audit of admitted nonpasses and excluded passes; the rates do not rank the pipelines.

\paragraph{Executable code benchmarks.}
HumanEval evaluates programs synthesized from docstrings with unit tests and introduced pass@$k$ reporting for repeated samples \citep{chen2021codex}. MBPP provides short Python problems with tests, while APPS evaluates a broader range of coding challenges using test cases \citep{austin2021mbpp,hendrycks2021apps}. These benchmarks establish execution-based functional evaluation as a practical methodology. Our study does not propose a competing benchmark; it audits which evidence survives when similar artifacts are considered for data construction.

AlphaCode uses large-scale sampling followed by filtering and clustering based in part on program behavior \citep{li2022alphacode}. That work demonstrates the importance of selection after generation. The present audit focuses on a preceding accounting issue: the selection predicate should be named according to what it observes, especially when selected code may later become supervision.

\paragraph{Test adequacy and generated tests.}
EvalPlus expands HumanEval tests and exposes failures missed by the original suite \citep{liu2023evalplus}. It strengthens evaluation; our audit tracks both admitted nonpasses and excluded passes under a fixed admission rule. CodeT uses generated tests to rank code candidates through execution agreement \citep{chen2022codet}, rather than each candidate's self-test completion alone. TCGBench evaluates valid test-input generators and targeted generators for bugs in human-written competition code \citep{cao2026tcgbench}. We evaluate neither a new test generator nor these selection methods.

S* revises parallel samples using public-test feedback, clusters candidates by execution outputs, and selects between clusters using adaptively generated distinguishing inputs and execution-grounded LLM judgments \citep{li2025sstar}. Its discussion of selection policies identifies errors in generated expected outputs on poorly chosen inputs as a source of selection noise \citep[Section~5.4]{li2025sstar}. Our contribution is not that possibility itself. We audit bidirectional admission changes under explicit execution contracts, connect them to task-level returns and abstentions, and enumerate paired order sensitivity in fixed saved pools. These contrasts evaluate neither S*'s selection accuracy nor its test-time compute gains.

\paragraph{Learning and revision from execution.}
CodeRL uses test outcomes in a reinforcement-learning framework and trains a critic to predict functional correctness signals \citep{le2022coderl}. Self-Debugging asks language models to revise their outputs with execution feedback or explanations \citep{chen2023selfdebug}. These approaches operate on feedback that is more informative than syntax alone. Our contribution is orthogonal: it separates completion, oracle agreement, and semantic conformance so that the reward or revision signal is not described more broadly than its evaluator supports.

\paragraph{Related evaluator measurements.}
Related measurement work separates duplicate-response preferences from target-selection utility in LLM judges \citep{usami2026darkcurrent}. We instead examine an execution pipeline: admission, independent-suite outcomes, and the candidate returned for each task. These are distinct evaluation objects, not replications of the same phenomenon.

\paragraph{Execution-filtered synthetic training data.}
Execution also selects material for code-model training. Haluptzok et al.\ generate programming puzzles and solutions, filter solutions with a Python interpreter, and fine-tune on the resulting synthetic pairs \citep{haluptzok2023selfteach}. SelfCodeAlign generates tasks from seed-code concepts, samples responses with tests, checks them in a sandbox, and selects passing examples for instruction tuning \citep{wei2024selfcodealign}. KodCode supplies a further synthetic-data pipeline with its own verification policy \citep{xu2025kodcode}. These pipelines motivate separating admission evidence from training value. We do not reproduce, rank or refute them: our saved-pool audit traces how a declared execution-contract change propagates from candidate admission to a fixed selector's task-level returns and abstentions.

\section{Evidence Gates and Observable Outcomes}
\label{sec:gates}

The audit separates a candidate's structural form, its completion under an execution contract, the health of the evaluating oracle, and conformance to the task specification. Table~\ref{tab:gates} gives their units. In the public-task study, $S$ is the structural screen, $E$ is isolated self-test completion, and $A=S\land E$ is admission; $A^*$ uses the restricted module-supply condition. Compilation alone, $C$, is retained as an ablation baseline. Section~\ref{sec:public} specifies these procedures.

\begin{table}[t]
\centering\small
\caption{Evidence records answer different questions. Candidate suite outcomes are recorded separately from oracle health.}
\label{tab:gates}
\begin{tabular}{@{}>{\raggedright\arraybackslash}p{0.10\linewidth}>{\raggedright\arraybackslash}p{0.23\linewidth}>{\raggedright\arraybackslash}p{0.57\linewidth}@{}}
\toprule
Record & Unit & Observed target \\
\midrule
$S$ & Candidate artifact & Compilation, entry-point form, and the fixed counted-assertion contract \\
$E$ & Candidate--environment pair & Completion of the preserved self-tests under the declared execution contract \\
$O$ & Oracle--task pair & Reference compatibility and discrimination on the stated health controls \\
$M$ & Candidate--specification pair & Conformance over the intended domain, supported here only in the declared finite synthetic control \\
\bottomrule
\end{tabular}
\end{table}

For suite $o$ and environment $e$, retain the reported state $Y(c,o,e)$ and define $P(c,o,e)=\mathbf{1}\{Y(c,o,e)=\mathrm{PASS}\}$. A nonpass indicator keeps its underlying fail, timeout, unavailable, or hold state. $P_b$ and $P_+$ report agreement with the pinned base and expanded suites; $O$ concerns the instrument, not an individual candidate. We write $O$ and $M$ for the oracle and semantic evidence denoted $G_O$ and $G_M$ in the historical definitions. The claim ledger in Section~\ref{sec:limitations} states the reach of these observations. Historical $G_S/G_E$ use different contracts and remain separate in Appendix~\ref{app:historical}.

\section{Public-Task Study: Design and Cohort Outcomes}
\label{sec:public}

RQ1 compares candidate-level admission with pinned-suite outcomes in both directions, pairing inline $A$ with restricted module-supply $A^*$. RQ2 uses implementation substitution, registration on/off, and assertion-local probes only in the selected original-cohort subset. RQ3 replays declared selectors at the task level, retaining passage, nonpass and abstention and enumerating all 24 common sample orders. These analyses have different units; none treats a candidate-retention gain as a returned-task gain.

\paragraph{Study identifiers.}
The original and disjoint follow-up cohorts are archived as V9 and V11, respectively; these labels appear in plots and saved data. V10 identifies a planned cohort stopped by reference-health checks before generation, with no candidate outcomes. These are archive IDs, not model or manuscript versions. Public notation omits the archived subscript 9: $C,S,E,A$ name the same frozen contracts stored as C9/S9/E9/A9 in both cohorts. Saved figure and field labels retain their archive names.

The historical candidates have no matched independent oracle. We therefore designed a separate public-task study; it does not impute correctness labels to the 580 earlier candidates. The outcome-blind manifest was locally frozen before generation. The official HumanEval source commit is \nolinkurl{6d43fb980f9fee3c892a914eda09951f772ad10d}; the EvalPlus checker is pinned to commit \nolinkurl{26d6d00bb1fd0fa37f39c99d5290da67891d1c5e}, with HumanEvalPlus v0.1.10. A fixed SHA256 namespace ranking selects 32 of 164 public tasks without candidate outcomes. Here, native suite execution uses the archived vendored adapter and native subset API, not a complete unmodified EvalPlus package. The vendor receipt records AST-exact ground-truth and correctness-check functions and byte-exact evaluation and special-oracle files. Its 424 base and 22,387 added inputs are checked against reference-derived outputs and task-specific comparators. All 32 references pass both suites, and all 32 explicit exception-raising negative controls are rejected. This establishes executable health on these checks.

\paragraph{Generation and independent measurement.}
Two service-reported profiles, Qwen3.8-Flash-Next and gpt-oss-120b, each receive four attempts per task, for 128 planned attempts per profile and 256 overall. Frozen request settings specify temperature 0.6 and a 4,096-token ceiling, with fixed profile-specific reasoning settings. They contain the public function specification and ask for code plus at least three self-tests, without canonical solutions, native test inputs, or expected outputs. Validation records retain format and termination outcomes, service-reported identity, and request/source hashes. These identities identify the recorded service profiles, not independently inspected model weights. Failed formats are retained in the planned denominator. No generation request is retried. Native independent tests are applied to every recoverable code module with valid syntax, including those whose self-tests fail or have invalid syntax.

The exact returned labels are \nolinkurl{qwen3.8-flash-next} (Flash) and \nolinkurl{openai/gpt-oss-120b}. Both request temperature 0.6, \nolinkurl{max_tokens=4096}, and non-streaming responses. Flash requests \nolinkurl{reasoning_effort=none} and template flags \nolinkurl{thinking=false}, \nolinkurl{enable_thinking=false}, \nolinkurl{reasoning_effort=none}; gpt-oss requests \nolinkurl{reasoning_effort=low}. These are recorded requests, not verified service-side enforcement. The retained UTC request-timestamp windows are 8 October 2026, 21:42:24--22:14:18 (Flash) and 22:14:23--22:39:05 (gpt-oss), and 9 October, 00:08:28--00:34:03 and 00:34:12--00:56:41, respectively. The supplementary metadata retain subsecond precision and source hashes; these are not response-completion times. The selected exports do not establish the historical deployment operator, external-service versus self-hosted status, backend version, quantization or weight digest. Current infrastructure and label prefixes are not substituted for those missing metadata. This reporting follows the methodological guidance of \citet{baltes2026llmguidelines}, not an additional journal submission requirement.

\paragraph{Notation and field mapping.}
The historical $G_S$ and $G_E$ are not changed. For the public-task study, $C$ denotes Python compilation of the code module, a syntax check without execution. $S$ requires both code and self-test modules to compile, a top-level definition of the named function, and at least three distinct top-level assertions calling it. Constant assertions, identical-expression comparisons, and duplicate assertion ASTs are excluded. These are operational screens, not assertions of independence, coverage, reachability, or correctness. $E$ denotes successful isolated self-test execution, and $A=S\land E$ is the new joint acceptance predicate. The frozen internal field named \texttt{G\_E} stores $A$, whereas \texttt{self\_exec\_status=PASS} stores $E$. They must not be pooled with the historical $G_E$ observations.

$P_b$ denotes passing all native base tests; $P_+$ denotes passing both base and added suites. Write $o_b$ for the pinned base suite, $o_{b+}$ for its base-and-added conjunction, and $e_\star$ for the frozen environment. Then $P_b=P(c,o_b,e_\star)$ and $P_+=P(c,o_{b+},e_\star)$. The frozen internal fields \texttt{G\_O\_base} and \texttt{G\_O\_plus} store these candidate test outcomes. They are distinct from $O(o,t)$, which concerns oracle health or adequacy. Neither finite public-suite pass is a general $M=1$ certificate. We retain the checker's reported pass, fail, top-level timeout, unavailable-syntax, and hold states. The pinned checker catches per-input exceptions, including time-limit exceptions, as fail; reported fail is not an exception-cause classification, and zero top-level timeouts does not imply absence of per-input deadline violations.

\paragraph{Two execution contracts.}
The original inline contract executes the preserved implementation and self-tests together. Because the generation prompt specifies neither module naming nor placement, the post-outcome restricted module-supply comparison supplies an explicit alternative: a callback registers only statically referenced absolute imports of \texttt{solution}, \texttt{your\_module}, or \texttt{total\_match\_module}, and uses an empty alias list otherwise. Other imports receive no adaptation. Define $E^*$ as valid format and successful self-test completion under this condition, and $A^*=S\land E^*$. Implementation and test bytes, the four-second self-test guard, and the original independent-suite labels remain fixed. This condition is not complete file/package loading: namespace copying leaves functions' original globals intact. Section~\ref{sec:fullcontract} gives the full-pool accounting and distinguishes this comparison from the selected same-batch registration control.

\paragraph{Isolation and analysis.}
Public-task candidates execute only in an existing private Docker image identified by its content ID, using Python 3.12.13. Network is disabled, the root is read-only, execution is non-root, capabilities are dropped, and new privileges are forbidden. One CPU, 8~GiB memory, 64 processes, and a 2~GiB temporary filesystem bound the batch. Native time limits use the pinned checker defaults: minimum four seconds and reference-runtime factor four. Self-tests have a four-second child limit and six-second parent join. The setup is not a security benchmark. Before generation, 24 runner, 22 analysis, 16 primitive, and nine actual controlled-container fixtures passed independent checks.

The fixed analysis reports $A$, $P_b$, $P_+$, $A\land\neg P_b$, $A\land\neg P_+$, and $A\land P_b\land\neg P_+$, retaining all planned attempts and both configurations. It also reports independent-test nonpass fractions conditional on $A$, with their denominators. All four samples within a task are resampled together in a 32-cluster, 20,000-draw deterministic bootstrap. Conditional intervals are held if the original denominator is zero or more than 5\% of bootstrap draws have zero denominator. Degenerate intervals are not unseen-risk guarantees. Each configuration needs 128 generation and 128 independent-evaluation terminals with identity and hash reconciliation; otherwise performance remains null and the configuration is held. No model ranking, new significance test, or natural-error prevalence is inferred. The generation and oracle stages each have a fixed 90-minute limit and no resume.

The saved bootstrap uses Python \texttt{random.Random} with an integer SHA256 seed from the cohort namespace, the literal separator \texttt{|analysis|}, and profile ID. Task IDs are sorted; each draw samples 32 clusters with replacement. Percentiles use type-7 linear interpolation at 0.025 and 0.975; undefined conditional draws are counted and omitted from the percentile list before applying the stated hold rule. All four saved conditional analyses have zero undefined draws. The accompanying metadata record all four exact seeds, namespaces, source and output hashes. Verification covers these saved bindings, not a re-execution of the bootstrap; the original analyzer interpreter build remains unverified.

\paragraph{Compute and reproduction.}
Original-cohort generation records 34,636 prompt and 79,284 completion tokens for Flash and 40,484 prompt and 50,687 completion tokens for gpt-oss. Follow-up counts are 32,568 and 65,076 for Flash and 38,764 and 46,132 for gpt-oss, with no missing usage fields in the retained records. These are generation counts, separate from drafting jobs. Aggregate reconstruction applies fixed predicates and selectors to saved labels. Evaluation re-execution additionally needs saved programs, self-tests, the pinned checker and isolation recipe; generation replication additionally needs frozen prompts and profile/service access. The supplementary artifact's table-to-file map links the supplied saved records, reconstruction commands, and result fields to the selected reported aggregates.

\paragraph{Language-model assistance.}
The research questions and objectives were defined by the authors. AI-assisted tools configured by the authors, including large language models deployed locally on author-managed GPU infrastructure, supported implementation review and refinement, manuscript preparation, and internal critique. Outputs from these tools were not treated as independent validation of the experimental results. The authors take full responsibility for the final manuscript.

\subsection{Complete outcomes and both directions of disagreement}
\label{sec:outcomes}

\paragraph{Execution contract and interpretation.}
These disagreements are measured under the declared original inline execution contract. Exclusion of a suite-passing attempt does not by itself identify a wrong assertion expectation: the prompt leaves module naming and placement unspecified, so self-test completion can depend on the supplied execution contract. Section~\ref{sec:refsub} diagnoses designated rejections, and Section~\ref{sec:fullcontract} compares alternative module supply over the saved pools. Those later results distinguish contract-sensitive completion from the separate question of assertion/reference disagreement; the original suite labels remain fixed.

Both profiles complete all 128 generation and 128 native-evaluation terminal records, with reconciled identities and hashes and no missing or ambiguous attempts. Table~\ref{tab:v9counts} retains the complete generation denominator. Flash has five invalid response formats; these are not recoverable code modules and are marked not evaluated by the independent checker, rather than interpreted as semantic errors. All 128 gpt-oss formats are valid. No top-level native oracle timeout is recorded. All 11 admitted expanded-suite nonpasses are reported added-suite failures after base-suite passes; their exception causes are not identified by that status alone.

\begin{table}[tb]\centering\caption{Complete attempt accounting for the original cohort. Each profile has 128 generation and independent-evaluation terminal records (32 tasks, four attempts each); terminal coverage does not mean that all 128 underwent native execution. Invalid formats remain not evaluated. Rows are overlapping candidate-level predicates, not distinct outcome categories or task-level returns. The admitted expanded-suite nonpass fractions use $A$ denominators: 4/80 for Flash and 7/100 for gpt-oss. Finite reference-relative suite outcomes are not semantic-defect rates or classifier-sensitivity estimates.}\label{tab:v9counts}\begin{tabular}{lrr}\toprule Observation & Flash & gpt-oss\\\midrule
Planned attempts &128&128\\
Format-valid modules & 123 & 128\\
$C$: code compilation & 123 & 128\\
$E$: self-test completion & 88 & 103\\
$A$: joint acceptance & 80 & 100\\
$P_b$: native base pass & 123 & 123\\
$P_+$: native base and added pass & 119 & 113\\
$A\land\neg P_b$ & 0 & 0\\
$A\land\neg P_+$ & 4 & 7\\
\bottomrule\end{tabular}\end{table}

The admission predicate accepts 80 Flash and 100 gpt-oss candidates. Of these, 4 and 7 fail the expanded suite: 4/80=5\% and 7/100=7\%. The frozen 20,000-draw task-cluster intervals are [0,16]\% and [0.93,16.84]\%, respectively. These intervals describe resampling of the fixed task cohort, not a population-risk certificate or a tested difference between profiles. There are no zero-denominator draws in either actual conditional analysis. All admitted candidates pass the base suite; the resulting degenerate empirical interval does not certify absence of unseen base-suite failures.

The opposite disagreement is also substantial within this cohort (Table~\ref{tab:fullcontract}). Inline $A$ excludes 43 Flash and 20 gpt-oss candidates that pass both independent suites; restricted module-supply $A^*$ excludes 23 and 20. Among the inline exclusions, self-test execution fails for 35 and 19 candidates; the remaining eight and one complete their self-tests but fail the structural requirement. Thus an excluded suite pass is not, by itself, evidence of an incorrect self-test expectation.

\begin{table}[tb]
\centering
\small
\caption{Full-pool joint admission under the inline contract ($A$) and restricted module supply ($A^*$). Each row retains 128 planned candidates and 32 tasks. N is admitted expanded-suite nonpass, Excl.~P is excluded suite pass, and First P/N/A separates returned pass, nonpass, and abstention. Uniform P is the exact expected passed-task count; decimals are displayed only as reading aids.}
\label{tab:fullcontract}
\begingroup\setlength{\tabcolsep}{3pt}
\setbox0=\hbox{\begin{tabular}{llrrrrr}
\toprule
Cohort/profile & Rule & \shortstack{Admitted\\attempts} & \shortstack{Admitted\\nonpasses (N)} & \shortstack{Excluded\\suite passes\\(Excl.~P)} & \shortstack{Returned tasks\\First P/N/A} & \shortstack{Uniform\\expected\\passed tasks}\\
\midrule
Original Flash & $A$ & 80 & 4 & 43 & 29/1/2 & $29$ \\
Original Flash & $A^*$ & 100 & 4 & 23 & 29/1/2 & $29$ \\
Original gpt-oss & $A$ & 100 & 7 & 20 & 24/3/5 & $49/2=24.50$ \\
Original gpt-oss & $A^*$ & 100 & 7 & 20 & 24/3/5 & $49/2=24.50$ \\
Follow-up Flash & $A$ & 83 & 4 & 34 & 29/2/1 & $59/2=29.50$ \\
Follow-up Flash & $A^*$ & 98 & 5 & 20 & 29/2/1 & $88/3\approx29.33$ \\
Follow-up gpt-oss & $A$ & 107 & 9 & 14 & 27/3/2 & $109/4=27.25$ \\
Follow-up gpt-oss & $A^*$ & 107 & 9 & 14 & 27/3/2 & $109/4=27.25$ \\
\bottomrule
\end{tabular}}
\typeout{AUDIT-TABLE-WIDTH: generated_full_contract.tex natural=\the\wd0 limit=\the\linewidth}
\ifdim\wd0>\linewidth\resizebox{\linewidth}{!}{\box0}\else\box0\fi
\endgroup
\end{table}

Invalid formats remain in the planned counts: original Flash/gpt-oss have five/zero and follow-up Flash/gpt-oss two/one. Seven actual diagnostic fixtures are excluded from these rows.

Table~\ref{tab:fullcontract} pairs inline $A$ and restricted module-supply $A^*$ on the same planned candidates and tasks. Module supply raises Flash admissions from 80 to 100 in the original cohort and from 83 to 98 in the follow-up, while gpt-oss counts and every fixed first-admitted suite outcome and abstention remain unchanged. Section~\ref{sec:fullcontract} explains why these admission gains need not increase the uniform expected passed-task count, including the original pool's saturation and the follow-up's added nonpass. The original cohort's four-cell partition is retained in Appendix~\ref{app:originalpublicpartition}.

\begin{figure}[tb]
\centering
\includegraphics[width=\linewidth]{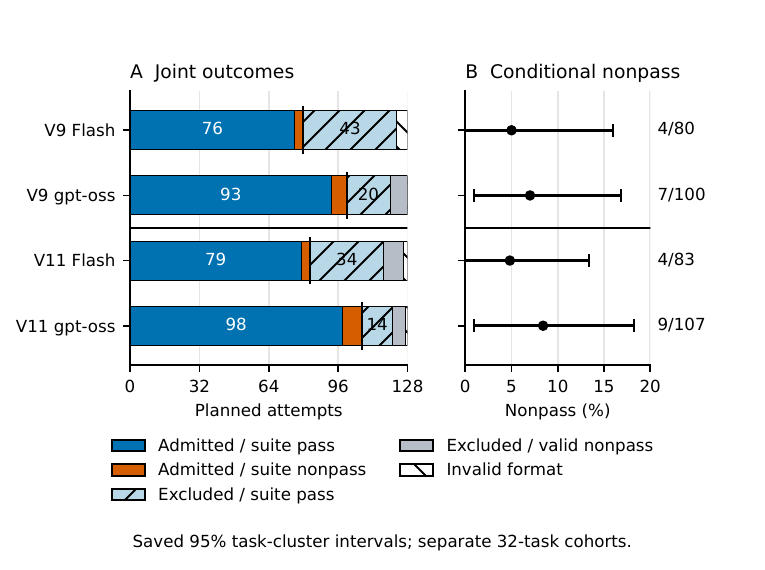}
\caption{Joint admission and expanded-suite outcomes in the separate original (V9) and disjoint follow-up (V11) cohorts. Each row retains all 128 planned attempts; invalid formats are displayed separately from valid nonpasses. The black tick marks the admitted total. Right: expanded-suite nonpass conditional on admission, with the existing frozen 95\% task-cluster intervals and exact numerator/denominator labels. The cohorts are not pooled, and neither suite outcomes nor intervals establish a general semantic-error rate or profile ranking.}
\label{fig:v9}
\end{figure}

\subsection{Task concentration and benchmark interpretation}
\label{sec:taskconcentration}

Flash's four admitted expanded-suite failures all concern \texttt{HumanEval/151}; gpt-oss's seven concern four tasks (151, 50, 29, and 43). The profiles contain 16 and four duplicate code rows, respectively. Samples from one task and repeated implementations are therefore not independent replications of an error mechanism. All 11 observed failures are indexed against the pinned added-input list, with code, self-test, and witness-input hashes.

A concrete example illustrates why the scope matters. On task 151, one Flash candidate explicitly accepts integer-valued floats and includes a self-test that counts \texttt{[1.0,3.0]} as contributing ten. The native reference uses a different type-sensitive rule; the recorded added-suite failure is on \texttt{[1.0,3.5,-4.6]}. Other failures on this task involve Boolean inputs. On task 50, one admitted gpt-oss candidate preserves nonalphabetic characters, whereas the provided encoder and native reference apply their modular character transformation to every character; its recorded failing input contains spaces. These are observed disagreements with the pinned benchmark contract. Questions about the intended input domain or a human reading of ``integer'' remain separate from the test outcomes. We do not relabel all such disagreements as human-verified semantic defects.

\subsection{Independent-suite discrimination controls}
\label{sec:discrimination}

Before inspecting the natural candidate outcomes, we separately committed 64 trusted controls on the same 32 tasks: a canonical implementation and an otherwise identical wrapper that raises an exception on one hash-selected added input absent from the base-input list. After the natural batch completed, one native batch evaluated these controls under the same checker and isolation settings. All 32 references pass both suites. All 32 selective-fault wrappers pass the base suite and fail the added suite, with each recorded failure index matching its precommitted witness. This verifies discrimination of those deliberately introduced violations. These controls are separate from the 256 generated candidates and the earlier 456 finite-domain programs.

\subsection{Reference-health hold before candidate generation}
\label{sec:refhold}

Figure~\ref{fig:cohortselection} summarizes the three cohort-selection stages and the pre-generation reference-health hold.

After observing the task concentration above, we selected a disjoint follow-up cohort of 32 tasks from the remaining 132 public tasks using a new fixed SHA256 namespace. This follow-up is outcome-motivated; it was not part of the original study's precommitment. Its protocol requires reference health before any new candidate generation and forbids replacing an unfavorable task.

Thirty-one pinned references pass both suites; all 32 negative controls are rejected. However, the \texttt{find\_zero} reference on \texttt{HumanEval/32} passes its 100 base inputs and fails seven of 788 added inputs under the native polynomial-residual criterion. For one failing input, its returned root is approximately 0.04912699 and the native residual is approximately $-142.99276$, compared with tolerance $10^{-4}$. A separate 100-digit Decimal Horner evaluation confirms tolerance violations for all seven recorded roots. The input lists have even coefficient counts and nonzero leading coefficients, as specified. 

The pre-generation kill rule holds this follow-up. No model requests are made, no task is replaced, and no tolerance or reference is silently repaired. These records are an instrument-health diagnostic, not another generated-code performance cohort. Task 32 is absent from the original 32-task study, whose manifest, generation and oracle records, and analysis remain unchanged. The diagnostic reinforces the need to check the pinned reference and comparison contract together rather than inheriting health from a benchmark name.

A subsequent, outcome-motivated health audit exhausts the 100 public tasks outside both 32-task cohorts, without rerunning the previously inspected 64. The same unmodified health checker, vendor sources, Python version, restricted image, and comparison contracts are used. All 100 canonical references fully pass both suites, and all 100 explicit exception controls are rejected. The resulting three-batch coverage ledger therefore records 163 healthy references and one held reference among 164 tasks. 

\begin{figure}[tb]
\centering\small
\begin{tabular}{@{}p{0.21\linewidth}p{0.70\linewidth}@{}}
\toprule
Public task set & Fixed selection and health sequence \\
\midrule
164 tasks & $\longrightarrow$ Original namespace: 32 tasks (V9), generation and suites \\
Remaining 132 & $\longrightarrow$ Second namespace: 32 tasks (V10), reference health first \\
V10 health & \texttt{HumanEval/32}: 7/788 added-input failures $\longrightarrow$ HOLD, zero generation \\
Remaining 100 & $\longrightarrow$ Third namespace: 32 tasks (V11), fixed before viewing the running health audit's outcomes and before V11 generation \\
100-task audit & 100/100 reference-health checks pass; V11 generation uses its fixed 32 tasks \\
\bottomrule
\end{tabular}
\caption{Cohort selection and the pre-generation health hold. V10 and V11 are motivated by earlier outcomes; each namespace is fixed without inspecting its new candidates. The V11 protocol fixes selection before viewing health outcomes, not before the health audit begins. No task is replaced after a health or candidate outcome. Archive IDs identify the three separate stages.}
\label{fig:cohortselection}
\end{figure}

\subsection{Completed disjoint generation follow-up}
\label{sec:v11}

Before viewing the complementary 100-task health outcomes, a new SHA256 namespace fixed 32 tasks from that set. These tasks are disjoint from both the original generation cohort and the held follow-up. All 32 pass the already recorded reference-health checks. The exact generation prompt, served-profile settings, parser, admission predicates, native sources, restricted image, and analysis contracts are retained. The new manifest and namespace are frozen before generation; reversible source adaptation and exact-message checks independently verify the changed task schedule. Existing native fixtures are reused as source evidence, not counted as newly executed controls. This extension is motivated by the earlier outcomes and is not part of the original precommitment. 

The selected cohort has 202 base and 26,206 added inputs. Each profile completes all 128 generation and 128 native-evaluation terminals, with no missing or ambiguous attempts. Flash has two invalid formats and gpt-oss one; all remain in the planned denominator. Table~\ref{tab:v11counts} reports this cohort separately. The same $A$, $P_b$, and $P_+$ symbols denote the unchanged predicates. No top-level native timeout is reported, with the per-input exception caveat above.

\begin{table}[tb]\centering\caption{Disjoint follow-up candidate outcomes, not pooled with the original cohort. Each profile retains 128 planned attempt slots from 32 tasks with four attempts each, and every slot has a terminal accounting record for generation and for independent evaluation. Invalid formats remain in accounting but are not native checker evaluations. Rows are overlapping predicates, not mutually exclusive categories or task-level returns. Conditional admitted expanded-suite nonpass uses 83 Flash and 107 gpt-oss $A$ admissions as denominators (4/83 and 9/107), not all 128 attempts. These finite reference-relative outcomes do not establish semantic-defect counts; task-level passage and abstention require the separately stated selection policy.}\label{tab:v11counts}\begin{tabular}{lrr}\toprule Observation & Flash & gpt-oss\\\midrule
Planned attempts & 128 & 128\\
Format-valid modules & 126 & 127\\
$C$: code compilation & 126 & 127\\
$E$: self-test completion & 89 & 109\\
$A$: joint acceptance & 83 & 107\\
$P_b$: native base pass & 120 & 123\\
$P_+$: native base and added pass & 113 & 112\\
$A\land\neg P_b$ & 0 & 0\\
$A\land\neg P_+$ & 4 & 9\\
\bottomrule\end{tabular}\end{table}

Expanded-suite nonpass conditional on admission is 4/83=4.82\% for Flash and 9/107=8.41\% for gpt-oss. Separate 20,000-draw task-cluster intervals are [0,13.33]\% and [0.95,18.27]\%; neither analysis has undefined conditional draws. All 13 admitted nonpasses pass the base suite and are reported as added-suite fail. Their unconditional frequencies are 4/128 and 9/128, with intervals [0,8.59]\% and [0.78,15.63]\%. 

For Flash, the four $(A,P_+)$ cells $(1,1),(1,0),(0,1),(0,0)$ are 79, 4, 34, and 11; for gpt-oss they are 98, 9, 14, and 7. Thus the same admission rule excludes 34 and 14 expanded-suite passes. Admitted nonpasses concentrate on two Flash tasks (\texttt{HumanEval/91}: three, \texttt{/76}: one) and four gpt-oss tasks (\texttt{/91}: four, \texttt{/76}: two, \texttt{/55}: one, \texttt{/163}: two). Duplicate-code counts are 16 and five. We retain this task concentration.

Reconciliation against the original journals covers all 256 identities, code/test hashes, served labels, denominators, and runtime records. The separately exported offline analyzer reproduces every original field except its explicitly verified source-hash substitution; eight corrupted-record/source fixtures are rejected before statistics. This offline reproduction makes no new model calls and leaves both cohorts' original records and analyses unchanged.

\paragraph{Answer to RQ1.}
The fixed rule disagrees with the pinned expanded suite in both directions in each cohort/profile. Across these four cells it admits 4--9 nonpasses per 80--107 admissions and excludes 14--43 suite passes per 128 planned attempts inline, or 14--23 under restricted module supply (Table~\ref{tab:fullcontract}). These are candidate-level observations, not returned-task effects or unique diagnoses of assertion error.

\section{Post-Outcome Contract and Selection Diagnostics}

Table~\ref{tab:fullcontract} pairs execution contracts and reports candidate admission beside task-level return. Table~\ref{tab:ablation} isolates candidate-set trade-offs of the fixed predicates; Table~\ref{tab:selectionreplay} reports returns and abstentions; Table~\ref{tab:ordersensitivity} tests dependence on sample order. The intervening rejection diagnostics answer RQ2 only for the selected original-cohort subset. Each diagnostic is frozen before its own execution or replay, not retroactively included in the original outcome-blind design.

\subsection{Interpreting self-test rejection}
\label{sec:refsub}

Reference substitution shows that self-test rejection persists for most of the selected original-cohort candidate attempts, even after replacing their implementations with pinned canonical references. The selection comprises all 54 attempts that pass the expanded native suite but fail self-test execution. These originally self-test-failing attempts form subsets of the $A=0,P_+=1$ cells in Table~\ref{tab:v9cross}: 35 Flash and 19 gpt-oss attempts. All 54 reference-substituted attempts pass the native base and added suites, yet under the preserved self-tests and fixed execution contract, self-test execution still fails for 34/35 Flash-origin and 19/19 gpt-oss-origin attempts (53/54 overall); one Flash-origin attempt passes. Reference substitution alone leaves unresolved alternatives: execution-contract incompatibility, including implementation-specific helpers or namespace assumptions; differing interpretations of the allowed input domain or task specification; and assertion/reference disagreement. Other execution conditions, nondeterminism, or environmental variation may affect the single changed outcome. We retain source, test and runtime identities for the subsequent contract-level diagnostics.

For the 23 Flash module failures, a same-batch control fixes code, tests and callback instrumentation while toggling module registration. Without registration, both implementation arms yield 23 \texttt{ModuleNotFoundError} outcomes; with registration, 22 originals and 21 references complete. This identifies registration-sensitive completion, one component of joint admission.

\paragraph{Unit-specific accounting.}
The 63 original-cohort excluded suite passes comprise 54 self-test failures (35 Flash and 19 gpt-oss) and nine self-test completions rejected by $S$ (eight and one). Table~\ref{tab:selfexceptions}(a) records the exception classes and the 23-candidate registration comparison; the latter does not recover completion for all 23 candidates. Table~\ref{tab:selfexceptions}(b) summarizes the candidate, assertion, and output-pair counts. Exception classes are observations, not established causes.

\begin{table}[tb]\centering
\caption{Selected original-cohort rejection diagnostics. (a) Inline exception classes for all 54 outcome-selected pairs, followed by restricted module-alias and same-batch on/off conditions on the 23 Flash ModuleNotFoundError cases. Diagnostic fixture runs (five inline, seven alias and seven controlled) are not included. (b) Unit-specific accounting of candidate, assertion, and output-pair observations. Code/test bytes and earlier outcomes are preserved; native suites are not rerun.}
\label{tab:selfexceptions}
\textbf{(a) Execution outcomes}\par\smallskip
\begin{tabular}{llrr}\toprule Condition/origin & Observed class & Original & Canonical\\\midrule
Inline Flash & ModuleNotFoundError & 23 & 23\\
Inline Flash & SystemExit & 2 & 2\\
Inline Flash & AssertionError & 10 & 9\\
Inline Flash & PASS & 0 & 1\\
Inline gpt-oss & AssertionError & 19 & 19\\\midrule
Alias Flash & AssertionError & 1 & 2\\
Alias Flash & PASS & 22 & 21\\
\midrule
Controlled off & ModuleNotFoundError & 23 & 23\\
Controlled on & AssertionError & 1 & 2\\
Controlled on & PASS & 22 & 21\\
\bottomrule\end{tabular}
\par\medskip
\textbf{(b) Unit-specific accounting}\par\smallskip
\begin{tabular}{@{}>{\raggedright\arraybackslash}p{0.17\linewidth}>{\raggedright\arraybackslash}p{0.75\linewidth}@{}}
\toprule
Unit & Selected observations \\
\midrule
Candidate & 54 inline originals: 23 ModuleNotFoundError, two SystemExit, 29 AssertionError. \\
Candidate & Same 54 substituted references: 23 ModuleNotFoundError, two SystemExit, 28 AssertionError, one completion. \\
Candidate & Registration on/off for 23 cases: off, 23 module failures in each arm; on, 22 original and 21 reference completions. \\
Assertion & 28 reference-AssertionError candidates: 129 assertions, 127 resolvable, 34 exact differences. \\
Output pair & Those 34 differences: three outside the pinned domain, two tolerance-compatible, 29 in-domain fresh-pair comparator disagreements. \\
\bottomrule
\end{tabular}
\end{table}

We separately examine the 19 gpt-oss and nine Flash reference-substitution \texttt{AssertionError} cases. Each resolved equality probes the pinned reference in a fresh namespace using exact Python \texttt{==}. Table~\ref{tab:expectationaudit} retains both selected subsets and unsupported assertions. Every selected case contains an exact discrepancy; all resolved probes return without exception or timeout.

Two examples make the expectation conflicts concrete. A \texttt{digitSum("woArBld")} test expects $213$, counting an uppercase \texttt{R} absent from the input; the docstring and reference give $131$. Two Flash \texttt{HumanEval/103} scripts each demand both \texttt{"0b1"} and \texttt{"0b10"} for \texttt{rounded\_avg(1,2)} through unconditional assertions, without function rebinding. These demands conflict for a deterministic function with unchanged behavior, independent of rounding convention.

A separate check of the 34 saved exact differences finds two compatible with the suite comparator's floating-point tolerance and three outside the pinned input contract. The remaining 29 in-domain output pairs still differ under a fresh comparator initialized with the task's tolerance. Two Flash scripts redefine the entrypoint, so the fresh probes do not reconstruct the original script's execution. Appendix~\ref{app:rejectiondiagnostics} gives the protocols, input-domain checks and comparator projection.

\paragraph{Answer to RQ2.}
In this original-cohort selected subset, reference substitution leaves 53 of 54 self-test rejections. The 23-candidate registration comparison yields 22 original and 21 reference completions when registration is on. Separately, 29 of 34 exact output differences remain in-domain disagreements under the declared fresh-pair comparator; these are output pairs, not 29 candidates or a mechanism prevalence estimate.

\subsection{Selected-input failure diagnostic}
After the follow-up is terminal, a frozen diagnostic selects all 13 admitted nonpasses and their first recorded added-suite failure inputs. It evaluates each original implementation and its pinned reference once on that input, without executing the co-generated self-tests. The diagnostic uses the recorded restricted image and an explicit 6~GiB process-memory guard, with a fixed four-second child limit and six-second parent join. Five actual fixture executions verify return-value, AssertionError, NameError, ValueError, and time-limit exception recording in the same 31-record batch. All 13 references return values. Ten candidates return different values, one raises ValueError, and two raise TimeoutException under this diagnostic limit. This fixed diagnostic limit is distinct from the native checker's reference-scaled per-input deadline; it does not retroactively classify every native fail or uniquely identify its cause.

Seven return-value disagreements occur on \texttt{/91} at the input \texttt{"I"}: candidates return one and references zero. Three occur on \texttt{/76} at $(-2,-2)$: candidates return false and references true.\footnote{Here ``reference'' denotes the pinned HumanEvalPlus/EvalPlus implementation used by this diagnostic, rather than an original HumanEval source listing.} On \texttt{/55}, the candidate raises ValueError at zero while its reference returns zero. Two \texttt{/163} candidates exceed the diagnostic limit at $(987654321,123456789)$, while their references return an empty list. These observations separate value disagreement, exception behavior, and time-limited computation at particular inputs. 

\subsection{Post-outcome admission-predicate ablation}
\label{sec:ablation}

To identify what the joint rule adds, we reconstruct five fixed predicates on every existing record: code compilation $C$, the structural requirement $S$, isolated self-test completion $E$, their joint rule $A$, and native base-suite passage $P_b$. This analysis is motivated after the cohort outcomes; it is not part of their original preregistration. The assertion threshold remains three, without threshold search, new generation, program execution, or resampling. Each cohort and profile retains its 128 planned attempts. Two separately implemented set-based and SQL aggregations agree on all 20 predicate cells and independently declared fixtures.

Table~\ref{tab:ablation} records admissions, nonpasses among admissions, suite passes excluded by each rule, and tasks retaining at least one suite pass. The last quantity is availability within the already generated four-sample pool; it is not the outcome of a deployed selection rule. Invalid formats stay in planned accounting. 

\begin{table}[tb]\centering\caption{Post-outcome fixed-predicate ablation. Every cohort/profile has 128 planned attempts and 32 tasks. $N$ is admitted candidates; $X$ admitted expanded-suite nonpasses; $L$ excluded expanded-suite passes; $T_+$ tasks with at least one admitted expanded-suite pass. The conditional nonpass denominator is $N$; no new interval or selection-utility claim is attached.}\label{tab:ablation}\begingroup\setlength{\tabcolsep}{3pt}
\setbox0=\hbox{\begin{tabular}{llrrrr}\toprule Cohort/profile & Predicate & \shortstack{Admitted\\attempts ($N$)} & \shortstack{Admitted\\nonpasses ($X$)} & \shortstack{Excluded\\suite passes ($L$)} & \shortstack{Tasks with\\suite pass ($T_+$)}\\\midrule
Original/Flash & $C$ & 123 & 4 & 0 & 31\\
 & $S$ & 111 & 4 & 12 & 31\\
 & $E$ & 88 & 4 & 35 & 29\\
 & $A$ & 80 & 4 & 43 & 29\\
\addlinespace
 & $P_b$ (reference) & 123 & 4 & 0 & 31\\
\midrule
Original/gpt-oss & $C$ & 128 & 15 & 0 & 30\\
 & $S$ & 120 & 12 & 5 & 29\\
 & $E$ & 103 & 9 & 19 & 26\\
 & $A$ & 100 & 7 & 20 & 25\\
\addlinespace
 & $P_b$ (reference) & 123 & 10 & 0 & 30\\
\midrule
Follow-up/Flash & $C$ & 126 & 13 & 0 & 30\\
 & $S$ & 118 & 11 & 6 & 30\\
 & $E$ & 89 & 4 & 28 & 30\\
 & $A$ & 83 & 4 & 34 & 30\\
\addlinespace
 & $P_b$ (reference) & 120 & 7 & 0 & 30\\
\midrule
Follow-up/gpt-oss & $C$ & 127 & 15 & 0 & 29\\
 & $S$ & 124 & 14 & 2 & 29\\
 & $E$ & 109 & 9 & 12 & 28\\
 & $A$ & 107 & 9 & 14 & 28\\
\addlinespace
 & $P_b$ (reference) & 123 & 11 & 0 & 29\\
\bottomrule\end{tabular}}
\typeout{AUDIT-TABLE-WIDTH: generated_admission_ablation.tex natural=\the\wd0 limit=\the\linewidth}
\ifdim\wd0>\linewidth\resizebox{\linewidth}{!}{\box0}\else\box0\fi
\endgroup\end{table}

Admission and selection answer different decisions. Relative to compilation-only $C$, joint $A$ removes 0, 8, 9, and 6 expanded-suite nonpasses but also excludes 43, 20, 34, and 14 expanded-suite passes, respectively, for original Flash, original gpt-oss, follow-up Flash, and follow-up gpt-oss. Among those removed nonpasses, 0, 5, 6, and 4 also fail the base suite, respectively. An admission decision therefore needs both excluded expanded-suite passes and admitted nonpasses reported per cell; the subsequent selection decision needs task-level passage and abstention. The replay below evaluates which retained candidate becomes the output.

Adding the structural rule to self-test completion reduces Flash admissions from 88 to 80 in the original cohort and from 89 to 83 in the disjoint follow-up. In both, admitted expanded-suite nonpasses remain four: the additional exclusions are eight and six suite passes. For gpt-oss, the original cohort loses two nonpasses and one suite pass (103 to 100 admissions), while the follow-up loses two suite passes and no nonpass (109 to 107). Original-cohort gpt-oss task availability decreases from 26 to 25; the other three structural additions do not change this task count. Thus this assertion-count requirement has cohort-dependent admission effects. Base-suite passage excludes no expanded-suite pass by construction, but still admits 4, 10, 7, and 11 expanded-suite nonpasses in the four separate cells. This fixed-predicate ablation does not implement CodeT's cross-sample execution-agreement selection \citep{chen2022codet}.

\subsection{Fixed-policy selection replay}
\label{sec:selectionreplay}

Availability of a passing candidate does not specify which candidate is selected. A further post-outcome replay therefore freezes one explicit policy for each of the five predicates: select the admitted candidate with the lowest original sample index, or abstain if none is admitted. The index, predicate, and tie rule do not inspect $P_+$. We also calculate the exact expectation for uniform selection among admitted candidates: a pool with $n$ admissions and $k$ expanded-suite passes contributes $k/n$ expected passes; an empty pool contributes one abstention. This is an analytic expectation, not an observed random draw. The $P_b$ policy uses recorded base-suite outcomes, which also contribute to the scored $P_+$ criterion; it is an oracle-informed retrospective reference, not a test-independent or cost-matched self-test baseline. 

Table~\ref{tab:selectionreplay} retains all 32 tasks per cell, separating passage, nonpass, and abstention. Reconstruction by a separately implemented SQL query against the original native journals agrees on all 20 cells and 640 task-level decisions. Selector fixtures verify rejection, numeric sample ordering, and row-order invariance; eligibility and index selection do not read the recorded $P_+$ field. The policy was specified after generation outcomes and is an offline replay within these four-candidate pools.

\begin{table}[t]
\centering\small
\caption{Fixed-policy saved-record selection replay. Each row contains 32 tasks. P/N/A denote first-choice expanded-suite pass, nonpass, and abstention. U is the exact expected number of passes under uniform selection among admissions; U/32 is expected task yield. Cohorts remain separate.}
\label{tab:selectionreplay}
\begingroup\setlength{\tabcolsep}{3pt}
\setbox0=\hbox{\begin{tabular}{lllrrrr}
\toprule
Cohort & Profile & Rule & \shortstack{Returned\\passes (P)} & \shortstack{Returned\\nonpasses (N)} & \shortstack{Abstained\\tasks (A)} & \shortstack{Uniform\\expected passed\\tasks (U)} \\
\midrule
Original & Flash & $C$ & 31 & 1 & 0 & 31 \\
Original & Flash & $S$ & 31 & 1 & 0 & 31 \\
Original & Flash & $E$ & 29 & 1 & 2 & 29 \\
Original & Flash & $A$ & 29 & 1 & 2 & 29 \\
\addlinespace
Original & Flash & $P_b$ (reference) & 31 & 1 & 0 & 31 \\
\midrule
Original & gpt-oss & $C$ & 28 & 4 & 0 & $113/4$ \\
Original & gpt-oss & $S$ & 27 & 4 & 1 & $109/4$ \\
Original & gpt-oss & $E$ & 25 & 3 & 4 & $51/2$ \\
Original & gpt-oss & $A$ & 24 & 3 & 5 & $49/2$ \\
\addlinespace
Original & gpt-oss & $P_b$ (reference) & 29 & 3 & 0 & $59/2$ \\
\midrule
Follow-up & Flash & $C$ & 29 & 3 & 0 & $115/4$ \\
Follow-up & Flash & $S$ & 29 & 3 & 0 & $173/6$ \\
Follow-up & Flash & $E$ & 29 & 2 & 1 & $59/2$ \\
Follow-up & Flash & $A$ & 29 & 2 & 1 & $59/2$ \\
\addlinespace
Follow-up & Flash & $P_b$ (reference) & 29 & 2 & 1 & $117/4$ \\
\midrule
Follow-up & gpt-oss & $C$ & 28 & 4 & 0 & $113/4$ \\
Follow-up & gpt-oss & $S$ & 28 & 4 & 0 & $113/4$ \\
Follow-up & gpt-oss & $E$ & 27 & 3 & 2 & $109/4$ \\
Follow-up & gpt-oss & $A$ & 27 & 3 & 2 & $109/4$ \\
\addlinespace
Follow-up & gpt-oss & $P_b$ (reference) & 28 & 3 & 1 & $113/4$ \\
\bottomrule
\end{tabular}}
\typeout{AUDIT-TABLE-WIDTH: generated_selection_replay.tex natural=\the\wd0 limit=\the\linewidth}
\ifdim\wd0>\linewidth\resizebox{\linewidth}{!}{\box0}\else\box0\fi
\endgroup
\end{table}

In the original cohort, first-choice expanded-suite passage changes from 31 to 29 Flash tasks and from 28 to 24 gpt-oss tasks when moving from $C$ to $A$, with two and five abstentions. In the separate follow-up, passage changes from 29 to 29 and from 28 to 27, with one and two abstentions. Relative to $C$, none of these four fixed-order joint-policy contrasts switches a selected nonpass to a pass: reduced selected nonpass counts arise through abstention. The additional structural requirement changes the original gpt-oss first-choice result from 25 under $E$ to 24 under $A$ by excluding a task whose selected candidate passed; the other three first-choice cells are unchanged. Uniform-choice expectations preserve the same distinction between coverage and passage. These observations support reporting the selection rule and its abstentions explicitly.

\subsection{Execution-contract comparison on all saved candidates}
\label{sec:fullcontract}

The full-pool comparison uses the two execution contracts defined in Section~\ref{sec:public}, extending beyond the selected import-failure diagnostic.

Table~\ref{tab:contractscopes} distinguishes the units and interventions. Reference substitution changes implementations in designated attempts; same-batch registration isolates registration-sensitive completion only in its designated import-failure subset (Section~\ref{sec:refsub}). The full-pool comparison below measures admission and task selection under alternative module supply, retaining the saved candidates and original suite labels. Assertion-local expectation and input/comparator checks (Appendix~\ref{app:rejectiondiagnostics}) are not additional candidate or task outcomes. 

\begin{table}[htbp]
\centering
\caption{What each execution comparison identifies. Completion remains separate from oracle adequacy and semantic correctness.}
\label{tab:contractscopes}
\small
\setlength{\tabcolsep}{4pt}
\renewcommand{\arraystretch}{1.12}
\begin{tabular}{@{}>{\raggedright\arraybackslash}p{0.22\linewidth}>{\raggedright\arraybackslash}p{0.33\linewidth}>{\raggedright\arraybackslash}p{0.39\linewidth}@{}}
\toprule
Comparison & Fixed and changed components & Supported reading \\
\midrule
Reference substitution & The same self-tests and execution condition; original implementation versus reference. & Completion with either implementation. A persistent rejection does not alone establish that a test is invalid. \\
\addlinespace[3pt]
Same-batch registration & Implementation, tests, callback, and driver fixed; declared aliases on versus off. & Registration-sensitive completion in the outcome-selected import-failure cases, under this instrumentation. \\
\addlinespace[3pt]
Full-pool contracts & Implementation and test bytes retained; earlier inline versus restricted namespace-alias module supply. Native-suite labels are retained. & Admission and selection under alternative contracts. Aliases do not provide full package loading; function globals are retained. Callback insertion and later execution prevent a registration-only causal interpretation. \\
\bottomrule
\end{tabular}
\end{table}

All 512 planned records remain fixed and the two cohorts remain separate. One bounded native batch executes the 504 format-valid candidates after seven actual fixtures match; the eight invalid records retain their planned denominators without new execution. Old $E$ and $A$ observations are retained. A separately implemented original-source SQL query reconciles all 16 admission and selection cells and 512 task-level decisions, alongside all 511 executed-record identities (504 format-valid candidates and seven fixtures). No model generation or native base/added-suite execution is repeated.

Table~\ref{tab:fullcontract} separates admission changes from the same fixed first-admitted policy and exact uniform-choice expectation. Flash joint admission increases from 80 to 100 in the original pool, adding 20 expanded-suite passes and no nonpass. In the follow-up it increases from 83 to 98, adding 14 passes and one nonpass. The corresponding execution-only admissions change from 88 to 110 and from 89 to 104. The gpt-oss execution-only and joint admissions remain unchanged in both pools. Every first-admitted suite outcome and abstention count is unchanged under both predicates, although the selected identity need not be unchanged. Flash's follow-up uniform expected passed tasks decreases from $59/2=29.50$ to $88/3\approx29.33$; the other joint expectations are unchanged. Expanding admission therefore does not improve the observed first-choice outcome and can admit an additional independent-suite nonpass within these saved pools.

Within this full-pool comparison, candidate-level records identify the two original Flash execution recoveries not admitted by $A^*$: sample 0 of \texttt{HumanEval/2} and \texttt{/81} completes under restricted module supply but has no counted distinct target assertion and hence fails $S$. Thus the 22 additional $E^*$ completions include 20 additional joint admissions; this relationship is checked by identity, not inferred from the aggregate totals alone. These full-pool record identities do not equate that intervention with the selected 23-case same-batch registration control.

\paragraph{Why 29 remains 29.}
For a nonempty admitted pool with $n$ candidates and $k$ expanded-suite passes, adding $m>0$ passes while retaining the old admissions changes its uniform expected pass contribution by
\[
\frac{k+m}{n+m}-\frac{k}{n}=\frac{m(n-k)}{n(n+m)}\geq 0,
\]
with equality exactly when $k=n$. Adding passes to an empty pool instead changes its contribution from zero to one; a task receiving no additions is unchanged. In the original Flash comparison, all 20 added passes enter already all-passing admitted pools, so the expected count remains $U=29$. This saturation concerns admitted pools, not all four planned samples. The two abstaining tasks, \texttt{HumanEval/123} and \texttt{/47}, retain five valid self-test failures and three invalid, unevaluated slots; all five valid candidates pass the independent suite. The follow-up also adds a nonpass, so the pass-only inequality does not apply to that contrast.

\paragraph{From results to reproduction artifacts.}
The supplementary artifact's table-to-file map links the supplied records, scripts, and result fields to the selected reported aggregates. Aggregate reconstruction from saved records does not re-execute the programs or independently validate the derivation of the recorded outcomes.

\subsection{Exact paired sample-order sensitivity}
\label{sec:ordersensitivity}
The numeric-index selector reports one order of four saved candidates. A further post-outcome replay enumerates all $4!=24$ sample-index orders, applying each order to both arms and every task within a cohort/profile. We compare $C\to A$ and $A\to A^*$ without generating candidates or executing code. Admission sets and suite outcomes are fixed before this replay and do not depend on order. Averaging all 24 permutations within a task makes each admitted index equally frequent as first choice; an empty set always abstains. Summing the task averages yields the uniform-choice expected passed-task count without assuming independence. Each common-order policy applies the same permutation of saved sample indices to all 32 tasks. The reported ranges cover these 24 policies, not the $24^{32}$ task-specific order combinations or prediction intervals for independent task randomization. Uniform-choice equivalence concerns each arm's expected pass count, not independence between the selections. Directional $N\to P$ and $P\to N$ expectations retain the shared within-task order; they do not describe independently randomized arm selections.

\begin{table}[htbp]
\centering\small
\caption{Exact sample-order sensitivity in the separate saved pools. Ranges cover all 24 global sample-index orders; means and expected $N\to P$ transitions average over those orders. $\Delta P$ is the new minus old number of selected expanded-suite passes among 32 tasks. Fractional entries are exact expectations, not observed counts.}
\label{tab:ordersensitivity}
\begin{tabular}{llrrrrr}
\toprule
 & & \multicolumn{3}{c}{$C\to A$} & \multicolumn{2}{c}{$A\to A^*$}\\
Cohort & Profile & $\Delta P$ range & Mean $\Delta P$ & Mean $N\to P$ & $\Delta P$ range & Mean $\Delta P$\\
\midrule
Original & Flash & $[-2,-2]$ & $-2$ & $0$ & $[0,0]$ & $0$\\
Original & gpt-oss & $[-5,-3]$ & $-15/4$ & $1/2$ & $[0,0]$ & $0$\\
Follow-up & Flash & $[0,2]$ & $3/4$ & $3/4$ & $[-1,0]$ & $-1/6$\\
Follow-up & gpt-oss & $[-1,-1]$ & $-1$ & $0$ & $[0,0]$ & $0$\\
\bottomrule
\end{tabular}
\end{table}

Table~\ref{tab:ordersensitivity} limits the earlier ``no selected nonpass becomes a pass'' observation to the numeric-index policy: other orders yield $N\to P$ transitions in original gpt-oss and follow-up Flash pools. Follow-up Flash's module-supply contrast yields a $P\to N$ transition in four of 24 orders, although the numeric order is unchanged. Its exact expected loss is $1/6$ of a passed task. The other three module-supply contrasts preserve every task's outcome in every order, even when selected identities change. A separately implemented SQL query reconstructs all 6,144 paired task/order decisions, 256 task histograms, and 192 order summaries directly from saved metadata; a separate earliest-union formula agrees for all 256 possible pairs of admission subsets. All eight numeric-order results and marginal expectations reconcile with the earlier frozen analyses.

\paragraph{Answer to RQ3.}
All four numeric-index $A\to A^*$ contrasts preserve returned suite outcomes and abstentions, despite the extra Flash admissions. In follow-up Flash, four common orders instead lose one passed task; the exact uniform expected count decreases from 29.50 to approximately 29.33. The result depends on the declared selector and placement of actual candidates, rather than admission totals alone.

\section{Implications for Evidence Audits}
\label{sec:implications}

\begin{table}[tb]
\centering\small
\caption{Research questions, answers and evidence. Candidate, assertion, output-pair and task-level units remain distinct.}
\label{tab:rqanswers}
\begin{tabular}{@{}>{\raggedright\arraybackslash}p{0.08\linewidth}>{\raggedright\arraybackslash}p{0.62\linewidth}>{\raggedright\arraybackslash}p{0.22\linewidth}@{}}
\toprule
RQ & Answer in the saved pools & Evidence \\
\midrule
RQ1 & Bidirectional disagreement: 4--9 admitted nonpasses per 80--107 admissions; excluded suite passes per 128 planned attempts range from 14--43 inline to 14--23 with restricted module supply. & Tables~\ref{tab:v9counts}, \ref{tab:fullcontract}, \ref{tab:v11counts}, \ref{tab:ablation}; Fig.~\ref{fig:v9} \\
RQ2 & Selected original cohort only: 53/54 rejections persist after reference substitution; registration on yields 22 original and 21 reference completions among 23 cases. Of 34 exact output differences, 29 remain in-domain fresh-pair disagreements. & Section~\ref{sec:refsub}; Tables~\ref{tab:selfexceptions}, \ref{tab:expectationaudit}, \ref{tab:comparatorpolicy} \\
RQ3 & All four fixed first-admitted outcome/abstention cells are unchanged by $A\to A^*$. Follow-up Flash loses one passed task in four of 24 common orders, with mean $-1/6$. & Tables~\ref{tab:fullcontract}, \ref{tab:selectionreplay}, \ref{tab:ordersensitivity} \\
\bottomrule
\end{tabular}
\end{table}

\subsection{For researchers: measure retention and return separately}
An admission rule changes which candidate records remain available; a selector changes which candidate becomes a task's output. The symmetric $A/A^*$ comparison measures both. Additional passing admissions can enrich a collection while leaving a fixed selector unchanged, as the original Flash pool illustrates. The follow-up's order sensitivity shows why reporting one numeric order alone can hide a selected-pass loss. Studies of generated-code collection should therefore report admitted nonpasses, excluded passes, returned outcomes, and abstentions as separate quantities.

Rejection diagnosis also needs its own unit. The selected registration control identifies completion that depends on module supply, whereas assertion-local probes compare particular expectations with a pinned reference and comparator. The full-pool comparison measures the combined contract change. Keeping these units distinct permits an explanation of the observed exclusions without assigning a single mechanism to every rejected candidate. CodeT's execution-agreement selection \citep{chen2022codet}, CodeRL's test feedback \citep{le2022coderl}, and Self-Debugging's revision loop \citep{chen2023selfdebug} use related signals for different decisions; their effectiveness is not evaluated by this audit.

\subsection{For practitioners: retain an evidence vector}
A practical evidence record follows five rules:
\begin{enumerate}
  \item Keep task, candidate and source identities with the structural decision and rejection reason.
  \item Record the execution contract and terminal state, preserving failure causes and unknown states.
  \item Link suite outcomes to the oracle version and its stated health controls.
  \item Report the selector, returned suite outcome and abstention separately from candidate admission.
  \item Use labels such as format repair, completed self-test or named-suite agreement; reserve stronger correctness labels for the domain supported by specification evidence.
\end{enumerate}

Data curation is one application of this record. A structural rejection may still be useful for repair, and an excluded suite pass may be useful supervision under a different collection objective. The present study measures neither training on those partitions nor their downstream value. Its contribution is the accounting needed to ask that question with a traceable execution and oracle contract.

\section{Threats to Validity}
\label{sec:limitations}

Table~\ref{tab:claimledger} collects the claim boundaries for the main observations; the following subsections distinguish population, measurement, inference and intervention limitations.

\begin{table}[tb]
\centering\small
\caption{Claim ledger for the main audit. Each observation retains its measured unit and scope.}
\label{tab:claimledger}
\begin{tabular}{@{}>{\raggedright\arraybackslash}p{0.24\linewidth}>{\raggedright\arraybackslash}p{0.66\linewidth}@{}}
\toprule
Observation & Scope of the claim \\
\midrule
Bidirectional disagreement & Candidate attempts in two separate 32-task cohorts under fixed predicates and pinned suites \\
Rejection mechanisms & Designated diagnostic subsets; reference disagreement is checked against input domains and comparators \\
Admission versus return & Four saved samples per task, fixed selectors and 24 common orders; costs and online generation are not evaluated \\
Reference health & Compatibility and targeted discrimination on the tested inputs; completeness and semantic truth remain separate \\
Data-construction uses & Proposed implications; downstream training effects are unmeasured \\
Supporting populations & Historical, oracle-diagnostic and finite-domain controls have distinct sources and units \\
\bottomrule
\end{tabular}
\end{table}

\subsection{External validity: study population and selection}
The main study covers 64 distinct HumanEvalPlus tasks and two served profiles, reported as separate cohorts rather than a model ranking. Namespace selection is not representative sampling of programming tasks. The follow-up and later diagnostics are motivated by earlier outcomes, with their own frozen procedures; they are not original precommitted arms. Four samples per task, repeated implementations, and high suite-pass rates restrict transfer to other budgets or less redundant pools. The cohorts use the same public benchmark rather than a fresh dataset or verified unseen pretraining items. Training contamination and served-weight identity remain unresolved.

This is a complete census of the saved candidate pools, not population estimation. Saturation explains an observed equality arithmetically; the empirical contribution is the traceable placement of actual admissions and their policy-dependent return consequences, not a novel saturation theorem or evidence of transfer to other filters.

Historical evidence has a different scope: one Python generator setting over 145 tasks, with no matched independent oracle. The 40-task oracle diagnostic includes a selected 20-task healthy subset and a 16-task resampling stage. The 456 planted programs span eight templates and declared finite domains. Their recorded rates, zero-discordance cases, and exhaustive finite checks retain those separate populations.

\subsection{Construct validity: measurement targets}
The three-assertion threshold is an operational format contract, not a validated test-quality score. Completion depends on the execution contract. Native suite outcomes are finite and reference-relative; type interpretation, input domains, and tolerances matter, as tasks 151, 91, and 76 illustrate. A checker-reported fail can include per-input time-limit exceptions, and the one-input probes use their own diagnostic deadline. The 64 selective-fault controls target known added inputs rather than a natural defect population. The held reference's failures do not establish that every input has a conveniently representable floating-point root or uniquely identify a numerical mechanism.

\subsection{Conclusion validity: aggregation and dependence}
Candidate disagreement clusters within tasks. The 20,000-draw task-cluster bootstrap recomputes conditional denominators in every resample; a zero original denominator or more than 5\% undefined draws holds an interval. These descriptive intervals do not identify population risk or a causal effect of assertion count, task properties, or resource limits. Exact uniform expectations are analytic quantities, not observed random selections. The contrasts share the four pre-generated candidates per task and the common sample orders defined in Section~\ref{sec:ordersensitivity}.

\subsection{Internal validity: execution contrasts and reproduction}
The module-supply comparison preserves candidate bytes and suite labels but changes the stated execution condition. Namespace copying is not complete file/package loading, and functions retain their original globals. Callback insertion and later execution distinguish this condition from the inline contract; the selected same-batch registration control supports only its narrower intervention, not a unique cause of every full-pool transition. Predicate costs are not equalized, and the replay does not measure sequential generation or screening costs under stop-at-first-admitted deployment. Generation, evaluation re-execution, and aggregate replay remain separate reproduction levels; no CodeT replication or downstream training effect is measured.

\paragraph{Overlapping study assets.}
The finite-domain programs and measurement fixtures overlap assets used in our separate study of LLM-judge reliability \citep{usami2026darkcurrent}. Here, those assets support the planted control experiment in Appendix~\ref{sec:controlled}, which reports execution and finite-domain validity outcomes; they do not supply the HumanEvalPlus generation outcomes. The other study measures judge responses to controlled comparisons. Shared assets are not independent replications. The public generation cohort and its admission analysis are separate from the judge-calibration calls.

\paragraph{Retained historical evidence.}
The historical 580-row receipt and three oracle-diagnostic files were hash-matched and their recorded counts reaggregated, without rerunning their generation or execution. Sanitized derivatives omit original output tails and operational metadata. A uniqueness check reconciles truncated task keys to full identities. Public redistribution remains subject to exact-asset approval and applicable rights.

\section{Conclusion}

A change of execution contract changes candidate admission without necessarily improving task-level returns or abstentions. In two separately frozen public-task cohorts, restricted module supply expands admission for one profile while all fixed first-admitted suite outcomes and abstentions remain unchanged. The same saved pools expose a selected-pass loss under four of 24 common orders in the follow-up. These are contract- and policy-conditional observations, not a ranking of filters or evidence of downstream training gains.

Original Flash's unchanged uniform expectation follows from added passes entering already all-passing admitted pools. In the follow-up, one additional admitted candidate fails the expanded suite, and the exact uniform-choice expectation decreases. Reference substitution and exception diagnostics explain why an isolated failure needs its execution contract before it can inform a test-validity claim.

The historical audit documents capture-ledger reconciliation; the planted finite examples delimit self-consistency claims on declared domains. Pre-generation reference-health checking also held a planned cohort when its pinned reference failed added-suite checks.

Future work should apply the same accounting protocol to other benchmarks and execution contracts. Rejection-mechanism diagnostics in the follow-up cohort remain future work. The downstream effects of training on retained and excluded partitions require direct measurement.

\section*{Statements and Declarations}
\paragraph{Data and code availability.}
The supplementary artifact provides typed outcome records for both cohorts, saved execution and diagnostic assertion/output-pair records, file-level hashes and retained record hash fields, table-to-file maps, and executable reconstruction scripts. These materials support selected saved-record aggregations, including the exception, registration, assertion, and comparator-policy diagnostics. The reconstruction does not rerun the models, candidate or reference programs, or bootstrap calculations.

The saved-record reconstruction artifact is publicly available at \url{https://github.com/usamihiroyasu/code-evidence-gates} (commit \nolinkurl{3425b76a8b091415fbe140eb845ea887203f61ca}).

\paragraph{Funding and competing interests.}
Computing infrastructure was supported through joint research funding from Mitsubishi Heavy Industries, Ltd., Research \& Innovation Center; Naohiko Matsuda is affiliated with this organization. The authors report no other competing interests.

\clearpage
\appendix
\section{Historical Audit: Scope and Summary}
\label{app:historical}
The earlier audit covers 145 tasks and four Python candidates per task under one recorded generator setting: 580 candidates, with no matched independent task oracle. Its structural gate counts at least three nonconstant assertion AST nodes. Its execution gate requires normal zero-return completion and a trusted marker under a digest-pinned container (no network, read-only root, non-root, one CPU, 1~GiB memory, 64 processes, 15-second candidate limit). These historical contracts differ from the public-cohort predicates.

The complete contingency cells $(G_S,G_E)=(1,1),(1,0),(0,1),(0,0)$ are 137, 392, 16 and 35. Thus $G_S$ accepts 529 and $G_E$ accepts 153 of 580. The task histogram for zero through four execution completions is 80, 19, 19, 12 and 15; return codes are 153 zero-with-marker, 426 code~1 and one code~$-9$. No recorded candidate timeout or infrastructure error occurs, and the kill's cause is not inferred. These are completion classifications, not correctness labels. Detailed definitions, tables and conditional rates are retained in the accompanying \emph{Detailed Supporting Methods and Historical Records}.

\section{Separate Oracle Diagnostic: Scope and Summary}
\label{sec:oracle}
A separate 40-task, two-arm diagnostic first executes each teacher-provided reference and its bundled self-tests. Twenty bundles complete. Both arms initially pass the same four tasks; none is discordant. For the remaining 16 healthy, joint-nonpass tasks, four additional samples per arm at temperature 0.7 recover no task (0/16). The health check establishes observed executability, not oracle completeness; zero discordance does not establish arm equivalence. Further recorded details of selection and generation are retained in the supporting companion, separate from both the 580 candidates and the public cohorts.

\clearpage
\section{Controlled Finite-Domain Experiment}
\label{sec:controlled}

The historical audit leaves semantic correctness unobserved. We therefore add a separate experiment in which the specification is decidable by exhaustive enumeration. This is a controlled demonstration of the gates, not additional sampling from the historical generator. Its units must not be pooled with the 580 candidates.

\paragraph{Tasks and ground truth.}
The suite contains eight templates with 12 parameter settings each: counting above a threshold, counting a specified value, a shifted sum, list rotation, clamping, parity of an affine sum, integer floor division, and closed-interval membership. Each item specifies a finite integer or integer-list input domain. A specification implementation independent of the candidate constructor checks all 51,960 input vectors and 155,880 candidate--input pairs. For each of 96 items, both a reference implementation and a syntactically different equivalent implementation agree with that specification on the entire domain; a deliberately faulty implementation disagrees on at least one input. These are finite-domain certificates, not claims about arbitrary Python inputs.

\paragraph{Interventions on the tests.}
Each program contains one of the fixed implementations and three assertions on distinct valid inputs. We form four complete groups of 96 programs: reference code with specification labels, equivalent code with specification labels, faulty code with labels taken from that same faulty code, and the same faulty code with specification labels including its first independent counterexample. A fifth group uses the faulty implementation with three specification-consistent inputs on which it happens to be correct. This group is possible for 72 items; 24 have fewer than three such inputs and are explicitly ineligible. Selection of the tests is deterministic and deliberately favors the corresponding diagnostic condition.

\paragraph{Execution and independent audit.}
All 456 programs are run once in Python \texttt{-I} subprocesses with a two-second limit and an external marker emitted after execution. This execution procedure is different from the historical Docker sandbox and is not a security evaluation. The programs are trusted synthetic templates. A separate post-hoc audit derives the assertion inputs and labels from the AST, checks the program identity and hashes, reevaluates specification conformance over the full domains, and reconciles the recorded exit code and marker. It does not reuse a model-generated verdict as ground truth.

\begin{table}[H]
\centering\small
\caption{Controlled programs. All satisfy $G_S$. Semantic validity is exhaustive on the declared finite domain. Test selection is planted; these counts do not estimate a generated-code error rate.}
\label{tab:controlled}
\begin{tabular}{@{}lrrr@{}}
\toprule
Implementation / tests & Programs & $G_E=1$ & $G_M=1$ \\
\midrule
Reference / specification labels & 96 & 96 & 96 \\
Equivalent / specification labels & 96 & 96 & 96 \\
Faulty / self-consistent labels & 96 & 96 & 0 \\
Faulty / independent witness & 96 & 0 & 0 \\
Faulty / weak specification-consistent tests & 72 & 72 & 0 \\
\midrule
Total & 456 & 360 & 192 \\
\bottomrule
\end{tabular}
\end{table}

\paragraph{What changes and what does not.}
For the paired 96 faulty implementations, changing the test labels and including an independent witness changes observed completion from 96/96 to 0/96 without changing the implementation. In the fifth group, even correct labels on three selected inputs permit all 72 faulty implementations to complete. Thus label independence and input discrimination are distinct requirements. Among the 360 completion-passing programs, 192 are valid and 168 invalid on the declared domains. The latter fraction is a construction-dependent count, not an estimated defect prevalence. The experiment gives a reproducible counterexample to using $G_S\land G_E$ as a correctness label and a concrete reason to record $G_O$ separately.

\section{Measurement Contracts and Denominator Conservation}
\label{sec:harness}
Two historical capture ledgers retain the same 580 candidate identities and code hashes but report 99 and 153 completions. A deterministic identity join recovers that difference. Allowing one trusted completion marker anywhere in a completed capture, rather than requiring final-byte placement, addresses an ordering-sensitive classification; repeated execution does not uniquely attribute every transition to stream order. The 54-record difference is not 54 additional correct programs.

A separate trusted termination fixture retains three primary attempts at a fixed fictional budget and one larger-budget diagnostic. Fixed-budget success is $1/3$ and success conditional on normal completion is $1/2$; the diagnostic cannot replace the primary record. All 24 row orders, three diagnostic variants and 12 malformed inputs were checked separately. These are accounting fixtures, not model observations. The supporting companion preserves the detailed marker, termination and parser contracts; missing or ambiguous observations remain holds, while known format failures remain in planned denominators.

\paragraph{Pinned native build check.}
\label{sec:publicbuild}
A separate release recipe pins a public \texttt{linux/arm64} Python 3.12.13 base by registry manifest digest, and NumPy 2.5.2 and psutil 7.0.0 wheels by content hashes. Inspection of the original content-addressed image identifies NumPy 2.5.2; the new public base has a different Python build timestamp. One build and three predeclared owned finite fixtures complete under the recorded isolation controls using the byte-identical native driver. The correct fixture passes both suites and self-tests; a self-consistent wrong fixture passes self-tests but fails both suites; a correct fixture with wrong self-tests fails that screen while passing both suites. These are new environment smoke checks, separate from all benchmark denominators. They establish that the release-candidate source/dependency recipe can run these controls on one arm64 host, not bitwise equivalence with the private image, other architectures, or a rerun of the original candidate studies.

\section{Self-Test Rejection Diagnostics}
\label{app:rejectiondiagnostics}

We conduct an outcome-motivated diagnostic on all 54 candidates that pass the expanded native suite but fail self-test execution: 35 from Flash and 19 from gpt-oss. Before new execution, a separate manifest commits this entire subset, the original code/test identities, and the pinned canonical implementation substituted for each original implementation. Every co-generated self-test byte is preserved. One bounded native batch uses the unchanged checker, driver, image and isolation settings; these transformed records are controls, not additional model generations or a fresh cohort.

All 54 substituted references still pass native base and added suites. Under the preserved self-tests, 34/35 Flash-origin cases and all 19 gpt-oss-origin cases fail; one Flash-origin case passes. Thus 53/54 observed self-test rejections persist after reference substitution. These observations isolate a compatibility question rather than automatically identifying wrong assertions or globally correct original code. Co-generated tests may rely on implementation-specific helpers, make a different input-domain assumption, disagree with the reference, or encounter other execution conditions. The one changed outcome also need not have a unique causal explanation in the presence of nondeterminism or environmental variation. We retain source/test/runtime hashes and these alternatives instead of treating every excluded native-suite pass as a proven semantic false rejection.

\paragraph{Bounded exception-class follow-up.}
A later manifest commits both original and canonical code for all 54 pairs, retaining the same self-tests, execution expression and namespace, four-second child limit, six-second join, memory guard, and restricted image. This diagnostic records only bounded exception-class names, without messages or traces; it does not rerun either native suite. One batch completes 108 paired-program records plus five actual PASS, AssertionError, NameError, ValueError, and TimeoutException fixtures, all of which match their declared outcomes. All original and canonical completion statuses agree with the previous records.

Table~\ref{tab:selfexceptions} separates the observed classes. The 35 Flash-origin originals yield 23 ModuleNotFoundError, two SystemExit, and ten AssertionError outcomes; the canonical arm retains the first two counts and yields nine AssertionError outcomes and one completion. All 19 gpt-oss-origin pairs yield AssertionError in both arms. Static inspection of the 23 module-failure cases finds 21 imports of \texttt{solution}, one of \texttt{your\_module}, and one of \texttt{total\_match\_module}. These are recorded integration and exception observations under the inline execution contract, not automatic semantic labels or unique explanations of the missing module. AssertionError denotes an observed assertion disagreement, and SystemExit denotes a caught exit request; neither by itself adjudicates the specification or intended success of a standalone script. No code, test, task, or primary outcome is repaired or relabeled in this follow-up.

\paragraph{Assertion-local expected-output audit.}
To localize the assertion disagreements, we freeze all 19 gpt-oss-origin reference-substitution failures and statically extract their 71 assertions. Sixty-nine direct equality assertions have resolvable literal inputs and expected values; two exception-control assertions remain unsupported. Each supported assertion calls the pinned reference in a fresh namespace, preserving Python's exact equality comparison, with a two-second CPU limit, a 256-MiB address-space limit and a three-second wall limit. The original sequential test scripts are not executed. Independent checks reconcile source bytes, assertion identities, resolved values and result counts.

Across 69 calls, 49 reference outputs equal the generated expectations and 20 differ; no call raises an exception or reaches the limit. Every selected case contains at least one discrepancy. For example, the generated expectation for \texttt{digitSum("woArBld")} is \texttt{ord('A')+ord('R')+ord('B')} ($213$), which includes an uppercase \texttt{R} absent from the input; the pinned docstring and reference give $131$. The median test uses six numbers, with central values 6 and 10 after sorting. Its generated expectation is 15, whereas their arithmetic mean is 8. These concrete value discrepancies complement the earlier exception classes.

For the \texttt{truncate\_number} assertion at input $123.000001$, the expected value is $10^{-6}$. The reference value is
\[
9.999999974752427\times10^{-7}.
\]
Exact floating-point equality fails. The rounding reference uses ties to even at $(2,3)$. These comparisons retain the assertion's literal equality rather than the native suite's task-specific comparator.

Static extraction resolves literals, containers, earlier literal bindings, restricted arithmetic and built-in \texttt{ord} in direct entrypoint equalities. It executes no imports, helper functions or other expected-side calls; unresolved assertions remain in the unsupported denominator.

\paragraph{Separate Flash-origin expectation audit.}
A later manifest applies the same assertion-local protocol to all nine Flash-origin reference-substitution \texttt{AssertionError} cases, spanning six tasks. All 58 retained assertions are resolvable direct equalities. One fresh, isolated reference call is made per assertion: 44 match and 14 differ, with no reference exception, timeout or unsupported form. All nine selected cases contain at least one discrepancy. Independent source checks and result reconciliation conserve every identity, argument, expected value and comparison. Both subsets are outcome-selected and retain within-task sample dependence.

The differences include three median expectations, one substring expectation, four grade-boundary expectations, two rounding expectations, three Collatz-list expectations and one exact floating-point comparison. For \texttt{truncate\_number(10.99)}, the generated expectation is $0.99$ and the pinned reference returns $0.9900000000000002$. This floating-point example belongs to \texttt{HumanEval/2}. Its script and the \texttt{HumanEval/7} script redefine their entrypoints, together accounting for ten retained assertions. Our protocol deliberately calls the pinned reference with the extracted literal arguments, rather than executing those redefinitions or replaying the original sequential script. An expectation discrepancy therefore does not by itself identify the first assertion that halted the original script.

A source-level inspection after the calls finds incompatible expected strings in both selected \texttt{/103} scripts: each contains equalities requiring \texttt{rounded\_avg(1,2)} to be both \texttt{"0b1"} and \texttt{"0b10"}. Both equalities are unconditional module-level assertions, with no function rebinding between them; neither script redefines the entrypoint. They impose incompatible demands on the same deterministic function with unchanged behavior, independent of rounding convention. In each script, one expected string matches the pinned output and the other differs. The inspection is post-outcome interpretation, not a prespecified population statistic or evidence that every rounding discrepancy is a specification error. All source bytes and comparison outputs are retained in the separate diagnostic ledger.

\paragraph{Comparator and input-domain sensitivity.}
We check all 34 saved exact differences against the pinned input contracts and a fresh-pair projection of the pinned comparison rule. In this hypothetical comparison, the generated expectation $e$ occupies the comparator's output argument and the saved reference value $r$ its expected argument. The finite scalar-float branch accepts
\begin{equation}
  |e-r| \leq a_t + 10^{-7}|r|,
  \label{eq:comparatorprojection}
\end{equation}
where, independently for each pair, $a_t$ is the task's declared tolerance, or $10^{-6}$ if that tolerance is zero and $r$ is a scalar float. The relative term and float-type trigger are anchored on $r$. After an exact mismatch, the nonzero-tolerance branch also requires matching output types. The saved reference values comprise nine finite scalar floats and 25 integers, strings, or integer/string lists; no float container or array occurs. \texttt{HumanEval/2} declares tolerance $10^{-6}$; the other eight selected tasks declare zero. All retained non-float pairs use exact equality, and no selected task uses the \texttt{find\_zero} special oracle. The projection ledger retains the exact \texttt{evalplus/eval/\_\_init\_\_.py} source and its hash. This fresh-pair projection does not reconstruct the mutable tolerance state or input order of the original suite. No candidate implementation, reference implementation, NumPy call, or native-evaluator call is made.

\begin{table}[htbp]
\centering\small
\caption{Policy projection of saved exact differences. The table groups rows outside the pinned input contract separately and reports comparator compatibility for in-domain rows. Counts are selected assertion rows, not independent candidates.}
\label{tab:comparatorpolicy}
\begin{tabular}{lrrrr}
\toprule
 & Exact & Outside & In-domain & In-domain\\
Profile & differences & contract & compatible & different\\
\midrule
gpt-oss & 20 & 3 & 1 & 16\\
Flash & 14 & 0 & 1 & 13\\
\bottomrule
\end{tabular}
\end{table}

Both scalar \texttt{HumanEval/2} examples are compatible with the floating-point tolerance. The three outside-contract gpt-oss rows comprise two \texttt{HumanEval/6} inputs containing characters not admitted by the pinned contract and one \texttt{HumanEval/129} $3\times3$ grid containing $10$ where the pinned contract requires a permutation of $1,\ldots,9$. The remaining 29 in-domain pairs differ under this comparator. Independent checks reconcile row identities and task contracts and verify the scalar inequality using a separately implemented decimal calculation. Table~\ref{tab:expectationaudit}'s exact-comparison tally is unchanged. Neither comparator disagreement nor membership outside this pinned contract establishes a general specification error; the subsets remain outcome-selected.

\paragraph{Restricted candidate-module alias condition.}
A separate post-outcome manifest fixes the 23 Flash-origin module-exception cases, spanning 16 tasks, and both implementation arms. It preserves each code and test byte, the original \texttt{\_\_main\_\_} namespace, and the four-second/six-GiB guard. The single composite execution inserts one reserved callback between implementation and tests; that callback copies implementation namespace bindings into a candidate-named module and registers only the statically imported \texttt{solution}, \texttt{your\_module}, or \texttt{total\_match\_module}. The callback binding and inserted statement are part of the intervention. Functions retain their original global namespace; subsequent module rebinding does not update those globals. This is an explicit namespace-alias condition, not a complete recreation of standalone file or package execution.

One new batch completes 46 paired records after seven declared fixtures pass: one completion for each of the three aliases, and the expected no-alias module failure, assertion, exit, and four-second timeout classes. Under the alias condition, 22/23 originals and 21/23 references complete; the remaining one and two records raise AssertionError. Twenty-one pairs complete in both arms, one original completes while its reference raises AssertionError, and one pair raises AssertionError in both arms. The earlier inline records remain unchanged. This condition sensitivity shows why reference substitution alone does not adjudicate test validity. The selected subset, later execution time, callback instrumentation, and limited module semantics constrain causal interpretation; neither native suite nor any model generation is rerun.

\paragraph{Same-batch registration control.}
To separate module registration from callback insertion and between-batch timing, a further frozen diagnostic evaluates all 23 cases under alias-on and alias-off conditions for both implementations. The byte-identical driver executes the same callback and composite expression in both conditions; only the declared alias list changes to an empty list. Hash-ordered four-record case blocks retain code/test bytes, image and resource limits. All 92 natural records and seven front-loaded actual fixtures complete, and independent original-journal SQL reconciles every identity and all 46 on/off pairs. Without registration, both arms yield 23 ModuleNotFoundError outcomes; with registration, 22 originals and 21 references complete, leaving one and two AssertionError outcomes. Every alias-on class agrees with the earlier alias batch. This paired intervention demonstrates registration-sensitive outcomes under the fixed instrumentation. It does not supply complete package semantics, adjudicate test validity or semantic correctness, or estimate a general failure rate. The outcome-selected subset and limited namespace semantics remain limitations; neither model generation nor either native suite is repeated.

\begin{table}[htbp]
\centering\small
\caption{Assertion-local expectations in separately selected original-cohort subsets. The subsets contain expanded-suite passes whose reference-substituted self-tests raise AssertionError. ``Unsupported'' retains exception-control assertions; every remaining assertion receives one fresh reference call using exact Python equality.}
\label{tab:expectationaudit}
\begin{tabular}{lrrrrrr}
\toprule
Profile & Cases & Tasks & Assertions & Match & Differ & Unsupported\\
\midrule
gpt-oss & 19 & 8 & 71 & 49 & 20 & 2\\
Flash & 9 & 6 & 58 & 44 & 14 & 0\\
\bottomrule
\end{tabular}
\end{table}

\section{Original Public-Cohort Partition}
\label{app:originalpublicpartition}
This table retains the original public cohort's four-cell inline partition. It is distinct from the historical 580-candidate audit. The symmetric inline/module-supply comparison appears in Table~\ref{tab:fullcontract}.
\begin{table}[tb]\centering\caption{Original-cohort candidate-level joint cells, with 128 attempt slots per profile (32 tasks, four attempts each). Unlike the overlapping predicate rows in Table~\ref{tab:v9counts}, these four cells partition attempts. The $A=1$ row pair contains 80 Flash and 100 gpt-oss admissions; only within those sets are nonpass fractions 4/80 and 7/100. The $A=0,P_+=0$ cell includes Flash's five invalid formats. These formats were not checker-evaluated and are not observed semantic failures. This cross-tabulation is not a task-return table or a classifier-sensitivity estimate.}\label{tab:v9cross}\begin{tabular}{lrr}\toprule Observation & Flash & gpt-oss\\\midrule
$A=1,P_+=1$ & 76 & 93\\
$A=1,P_+=0$ & 4 & 7\\
$A=0,P_+=1$ & 43 & 20\\
$A=0,P_+=0$ & 5 & 8\\
\bottomrule\end{tabular}\end{table}

\clearpage
\bibliographystyle{spbasic}
\bibliography{references}

\end{document}